\documentclass{llncs}

\usepackage{cmap}
\usepackage[utf8]{inputenc}
\usepackage[T1]{fontenc}
\usepackage[english]{babel}
\usepackage{verbatim}
\usepackage{hyphenat}
\usepackage{graphicx}
\graphicspath{{./images/}}
\usepackage[table,dvipsnames]{xcolor}
\usepackage{xspace}
\usepackage{array}
\usepackage{hhline}
\usepackage[caption=false]{subfig}
\usepackage{cleveref}
\usepackage{booktabs}
\usepackage{multirow}
\usepackage{tikz}
\usepackage{url}
\usepackage{cite}

\newcommand{\class}[1]{\textbf{\nohyphens #1}\xspace}
\newcommand{\attribute}[1]{\texttt{#1}\xspace}
\newcommand{\sstl}{\multicolumn{1}{c}{}}
\newcommand{\sst}[1]{
  \if A#1
    \multicolumn{1}{c|}{\cellcolor{black!10!white}#1}
  \else
    \multicolumn{1}{c|}{\cellcolor{black!10!white}#1}
  \fi
}
\newcommand{\ssl}[1]{\cellcolor{black!10!white}#1}
\newcommand{\cellLeftAligned}[1]{\multicolumn{1}{l|}{#1}}
\newcommand{\cc}[1]{\multicolumn{1}{c}{#1}}
\newcommand{\threelines}{\begin{tabular}{@{\hskip0pt}p{0pt}@{\hskip0pt}}\arrayrulecolor{white}\\\hline\hline\\\hline\hline\\\end{tabular}}

\colorlet{classInventory}{yellow!30!white}
\colorlet{classYear}{orange!50!white}

\begin{document}

\title{Tabula: A Language to Model Spreadsheet Tables}
\author{Jorge Mendes \and João Saraiva}
\institute{HASLab, INESC TEC and Universidade do Minho, Portugal\\
\email{\{jorgemendes,saraiva\}@di.uminho.pt}}

\maketitle

\begin{abstract}
  Spreadsheets provide a flexible and easy to use software development
  environment, but that leads to error proneness. Work has been done to prevent
  errors in spreadsheets, including using models to specify distinct parts of a
  spreadsheet as it is done with model-driven software development. Previous
  model languages for spreadsheets offer a limited expressiveness, and cannot
  model several features present in most real world spreadsheets.

  In this paper, the modeling language Tabula is introduced. It extends previous
  spreadsheet models with features like type constraints and nested classes with
  repetitions.  Tabula is not only more expressive than other models but it can
  also be extended with more features. Moreover, Tabula includes a bidirectional
  transformation engine that guarantees synchronization after an update either
  in the model or spreadsheet.

\end{abstract}

\section{Introduction}
\label{sec:intro}
%
%
The size and complexity of software has been quickly increasing in the
last years. A paradigmatic example is the size of the software
included in the aerospace industries: while the space shuttle
developed in the 80's contained about $400,000$ lines of source
code\footnote{\url{https://www.nasa.gov/mission_pages/shuttle/flyout/flyfeature_shuttlecomputers.html}},
the modern \textit{Airbus A380} built in this century includes more
than $100$ millions lines of code~\cite{wiels:hal-01184099}. In such
large and complex systems it is unfeasible to reason about the
software just by looking/understanding its source 
code~\cite{wiels:hal-01184099}.

Model-Driven Software Development (MDSD) has emerged as an important
software engineering discipline allowing developers to reason about
complex software by providing simple/concise abstractions - the
software model. Very much like a civil engineer develops
``human-scale'' models of a bridge, before the real bridge is
constructed, in MDSD a software engineer reason about his complex
software by analyzing a simpler model.

Spreadsheets are no exception, and, indeed, they tend to evolve into
large and complex software systems, which are difficult to understand,
to maintain, and to evolve. The combination of complexity with the
lack of abstraction mechanisms, is the main cause of the (too) many
errors caused by spreadsheets~\cite{Panko2000,panko06,rogoff}.

Model-driven spreadsheet development was introduced in
spreadsheets~\cite{Paine1997,Erwig2005,Engels2005,vlhcc2011} with two
main goals: Firstly, to provide a powerful abstraction of the business
logic of spreadsheets so that users can reason about their
spreadsheets by analyzing simple models, instead of very large and
complex data. Secondly, to provide a \textit{type system} for
spreadsheets: the model is incorporated into a regular spreadsheet so
that it limits the data/formulas that can be defined in the
spreadsheet cells. In such a model-driven spreadsheet engineering
setting, the spreadsheet data (i.e., the instance) has always to conform
to the spreadsheet model (i.e., the type), thus steering users in
introducing correct data.

ClassSheets were introduced by Engels and Erwig as a powerful
domain-specific modeling language for
spreadsheets~\cite{Engels2005}. ClassSheets define both the
computation and layout of spreadsheet tables. In previous work we have
extended the ClassSheet formalism to improve its
expressiveness~\cite{vlhcc2012}. Moreover, we conducted empirical
studies using real-world model-driven spreadsheets that showed an
improvement in users’ performance, while reducing the error
rate~\cite{tse2015}.  This latter work also showed the limitations of
(extended) ClassSheets: the business logic and layout of several
real-world spreadsheets could not be modeled by a ClassSheet. In other
cases, the ClassSheet model was not the natural way to express the
layout/logic of the spreadsheet.

In this paper we present a new modeling language for spreadsheets
tables: Tabula. This language is inspired by the ClassSheet modeling
language, namely its visual notation, but it provides more expressive
features like type constraints and nested classes with repetitions
enabled by a different abstract representation.
Moreover, Tabula includes a bidirectional transformation engine that
guarantees synchronization after an update either in a Tabula model or
spreadsheet.
Much like our previous work on ClassSheets, the usage of model-driven
spreadsheets targets repeated use of spreadsheets with a well-defined structure
and logic. Tabulae are to be defined by a specialist in the domain with
knowledge on modeling with Tabula, but usage of the respective spreadsheets
targets usual spreadsheet users.
We used the Tabula visual language to model a widely used
budget spreadsheet that is provided by Microsoft as a budget
template. Moreover, we also model the business logic of that
spreadsheet using a ClassSheet model, and we compare the expressiveness
of both spreadsheet modeling languages.

This paper is organized as follows: Section~\ref{sec:mdsd} gives a
short introduction to model-driven spreadsheets and discusses
the ClassSheet modeling
language. Section~\ref{sec:tabula} introduces in detail the Tabula
modeling language and briefly describes its bidirectional
transformation engine. In Section~\ref{sec:eval}, we evaluate the
expressiveness of Tabula when modeling a budget spreadsheet. We also
compare the Tabula and the ClassSheet models for this spreadsheet
instance. Finally, Section~\ref{sec:conclusion} includes our conclusions.

\section{Model-Driven Spreadsheet Development}
\label{sec:mdsd}
Before we propose a new modeling language for spreadsheets let us
discuss in more detail the state of the art on model-driven
spreadsheet engineering.

Several different approaches to use MDE in spreadsheets have been
proposed in literature. The first model-driven spreadsheet
specification language was \textit{Model Master}~\cite{Paine1997}: an
object-oriented textual specification of a spreadsheet that can be
compiled into a concrete spreadsheet and also to
be \textit{decompiled} from a spreadsheet. It has a mathematical
background, conceived from category theory concepts.

\textit{Spreadsheet templates}~\cite{Erwig2005} were the first approach to
define a MDE spreadsheet language with a visual representation which
allows to specify spreadsheets in a spreadsheet-like manner.  On top
of these spreadsheet templates, Engels and Erwig proposed
ClassSheets~\cite{Engels2005}: a high-level, object-oriented formalism
to specify the layout and business logic of spreadsheets. The visual
representation of ClassSheets is close to what spreadsheets users are
familiar with. ClassSheets are supported by their own application and
then compiled to spreadsheets. As a result, model and instance are
supported by different software systems, limiting the synchronization
whenever one artifact (model or instance) evolves. In order to
minimize this drawback, ClassSheets have been embedded in spreadsheet
system~\cite{vlhcc2011,tse2015}, bringing spreadsheet modeling closer
to end users, and allowing model/instance co-evolution. Moreover,
these embedded ClassSheets have been extended with additional features
to improve their expressiveness~\cite{vlhcc2012}.

\begin{figure}[ht]
  \centering
  \begin{minipage}[b]{.45\textwidth}
    \centering
    Spreadsheet\\[.5em]
    {\small\sf
    \begin{tabular}{|c|@{\hskip0pt}p{1\cmidrulesep}@{\hskip0pt}|l|r|}
      \hhline{~~|--}
      \sstl   & & \sst{A}        & \sst{B} \\\hhline{-~:==}
      \ssl{1} & & \textbf{Items} &         \\\hhline{-~|--}
      \ssl{2} & &         apple  &     5   \\\hhline{-~|--}
      \ssl{3} & &         banana &     2   \\\hhline{-~|--}
      \ssl{4} & &         cherry &     8   \\\hhline{-~|--}
      \ssl{5} & & \textbf{Total} & =SUM(B2:B4) \\\hhline{-~|--}
    \end{tabular}
    }
    \caption{Spreadsheet storing a list of items with their respective stock and
    its total.}
    \label{fig:example-items-spreadsheet}
  \end{minipage}
  \quad
  \begin{minipage}[b]{.49\textwidth}
    \centering
    \emph{Items} ClassSheet\\[.5em]
    \includegraphics[scale=0.3]{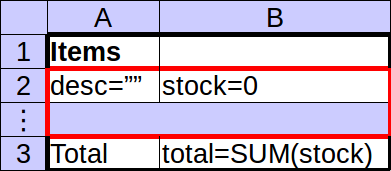}\\[.5em]
    \caption{ClassSheet specifying the spreadsheet in
    Fig.~\ref{fig:example-items-spreadsheet}, using the notation
  from~\cite{Engels2005}.\newline}
    \label{fig:example-items-classsheet}
  \end{minipage}
\end{figure}

To show the expressiveness of ClassSheets, and also their limitations,
let us consider a simple spreadsheet to keep track of the inventory of
products, as shown in Fig.~\ref{fig:example-items-spreadsheet}. In
its simplest form, the inventory is just a list of items. Each item
defines the name of the product (column~$A$) and the available
quantity (column~$B$). The last row contains the total of
products: the sum of the quantities of listed
products. Figure~\ref{fig:example-items-classsheet} contains the
ClassSheet (in its visual notation~\cite{Engels2005}) that models the
business logic and layout of this spreadsheet: it consists of a class,
named \class{Items}, that includes two attributes (\attribute{desc} of
type \emph{string}, and \attribute{stock} of type \emph{number}, with default values
the empty string and $0$, respectively). The total of products is
defined by a formula that refers to attribute names (and not the usual
column/row references). Finally, the layout is specified by the
vertical dots, meaning that values of the above attributes can repeat
vertically.

In the context of MDE we say that the spreadsheet instance presented
in Fig.~\ref{fig:example-items-spreadsheet} conforms to the
spreadsheet model in Fig.~\ref{fig:example-items-classsheet}. We can
also say that the spreadsheet data
in Fig.~\ref{fig:example-items-spreadsheet} has type \emph{Items} ClassSheet.

\begin{figure}[tb]
  \centering
  \small\sf
  \begin{minipage}[b]{.45\textwidth}
    \centering
    \begin{tabular}{|c|@{\hskip0pt}p{1\cmidrulesep}@{\hskip0pt}|l|r|}
      \hhline{~~|--}
      \sstl    & & \sst{A}        & \sst{B} \\\hhline{-~:==}
      \ssl{1}  & & \textbf{Inventory} &     \\\hhline{-~|--}
      \ssl{2}  & &                &         \\\hhline{-~|--}
      \ssl{3}  & & Fruit          &         \\\hhline{-~|--}
      \ssl{4}  & & apple          &     5   \\\hhline{-~|--}
      \ssl{5}  & & banana         &     2   \\\hhline{-~|--}
      \ssl{6}  & & cherry         &     8   \\\hhline{-~|--}
      \ssl{7}  & &                &         \\\hhline{-~|--}
      \ssl{8}  & & Legumes        &         \\\hhline{-~|--}
      \ssl{9}  & & beans          &     7   \\\hhline{-~|--}
      \ssl{10} & & peas           &    10   \\\hhline{-~|--}
      \ssl{11} & &                &         \\\hhline{-~|--}
      \ssl{12} & & \textbf{Total} & =SUM(B4:B6,B9:B10) \\\hhline{-~|--}
    \end{tabular}
    \caption{Spreadsheet to keep an inventory of items grouped in categories.}
    \label{fig:example-inventory-spreadsheet}
  \end{minipage}
  \quad
  \begin{minipage}[b]{.45\textwidth}
    \centering
    \includegraphics[scale=0.3]{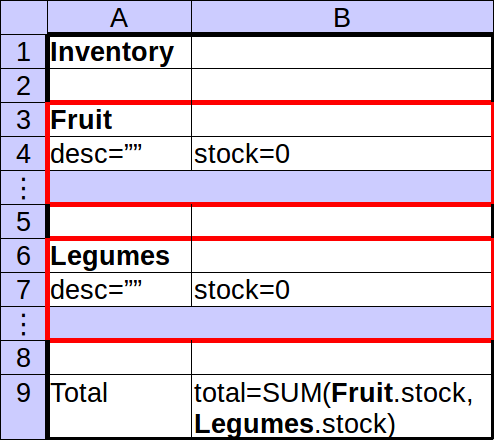}\\[.5em]
    \caption{ClassSheet specifying the spreadsheet in
    Fig~\ref{fig:example-inventory-spreadsheet}.}
    \label{fig:example-inventory-classsheet}
  \end{minipage}
\end{figure}

Let us consider now that we wish to distinguish different categories
of products, namely, \textit{fruits} and
\textit{legumes}. Figure~\ref{fig:example-inventory-spreadsheet} presents a
possible spreadsheet that structures the inventory in this
way. However, this new inventory spreadsheet is not easily specified
using ClassSheets, because they do not support nested repetition of
classes. Figure~\ref{fig:example-inventory-classsheet} contains a
ClassSheet that does model the new inventory, but, as we can see, some
of the data (the categories) must be defined at the model level. As a
result, models tend to grow and resemble the instance, which limits
the expressiveness of the approach: when a new product category has to
be considered both the model and instance have to be updated!

The key feature lacking in the ClassSheet modeling language is the
possibility to define nested repeated classes/attributes. Thus, in
ClassSheets, it is not possible to express the fact that we have
categories of products that repeat vertically, each of which has a
list of items (pairs of product name and quantity) that also repeat
vertically.

\begin{figure}[bt]
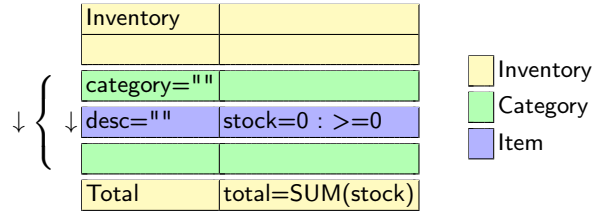

  \centering
  \small\sf
  \mbox{
  \begin{tabular}{r|l|l|}
    \hhline{~|-|-|}
    & \cellcolor{yellow!30!white}Inventory
      & \cellcolor{yellow!30!white} \\\hhline{~|-|-|}
    & \cellcolor{yellow!30!white}
      & \cellcolor{yellow!30!white} \\\hhline{~:=:=:}
    \multirow{3}{*}{$\downarrow \left\{ \threelines \right. \downarrow$}
      & \cellcolor{green!30!white}category=""
      & \cellcolor{green!30!white} \\\hhline{~:=:=:}
    & \cellcolor{blue!30!white}desc=""
      & \cellcolor{blue!30!white}stock=0 : >=0\\\hhline{~:=:=:}
    & \cellcolor{green!30!white}
      & \cellcolor{green!30!white} \\\hhline{~:=:=:}
    & \cellcolor{yellow!30!white}Total
      & \cellcolor{yellow!30!white}total=SUM(stock) \\\hhline{~|-|-}
  \end{tabular}
  }
  \quad
  \mbox{
  \begin{tabular}{|c|l}
    \hhline{-~}
    \cellcolor{yellow!30!white}~~ & Inventory \\
    \hhline{=~}
    \cellcolor{green!30!white}~~ & Category \\
    \hhline{=~}
    \cellcolor{blue!30!white}~~ & Item \\
    \hhline{-~}
  \end{tabular}
  }
  \caption{Tabula specifying the inventory spreadsheet in
  Fig.~\ref{fig:example-inventory-spreadsheet}.}
  \label{fig:example-inventory-tabula}
\end{figure}

The inventory spreadsheet can be modeled using Tabula as shown in
Fig.~\ref{fig:example-inventory-tabula}. In this case, categories can
also be abstracted in the model and all the data is defined in the
spreadsheet, only. A key advantage of Tabula over ClassSheets is the
ability to describe nested repetitions. In our running example, it
defines one for the category (as indicated by the downwards arrow on
the left of the left brace) and a nested one for the category items is
indicated by the other downwards arrow (on the right of the left
brace). This feature is not only possible for vertical repetitions
like in this example, but also for horizontal repetitions.  Moreover,
Tabula allows to define type constraints leading to a better
characterization of the data the spreadsheet should hold, and, as a
consequence it prevents user errors.  In our example, the
attribute \attribute{stock} has its values restricted to non-negative
numbers. A detailed description of the Tabula modeling language is
presented in the next section.

\section{The Tabula Spreadsheet Modeling Language}
\label{sec:tabula}
A spreadsheet comprehends both the logic and the layout of a
program. The Tabula modeling language specifies both of those
elements, having a definition of the computation a spreadsheet has to
perform, and how such computation is shown to users. Tabula include
also a bidirectional evolution engine that guarantees model/instance
synchronization after a transformation in one of these software
artifacts. In the next section we define the Tabula modeling
language. After that, we briefly describe the Tabula evolution engine.

\subsection{Specification of Layout and Logic}

A \emph{Tabula} is defined as a type with a name, a list of classes
and a grid. We express it in the Haskell programming language by the
following abstract data type:
\begin{verbatim}
data Tabula = Tabula Name [Class] (Grid TCell)
\end{verbatim}
Classes provide a meta-information about the cells in their range, while the
grid defines the layout and contents of the spreadsheet. For that, classes are
defined as
\begin{verbatim}
data Class = Class Name Range Expansion
\end{verbatim}
The range of type \verb|Range| is defined as the pair of its top-left coordinate and its
bottom-right one.
\begin{verbatim}
type Range = (Point, Point)
\end{verbatim}
The expansion information, of type \verb|Expansion|, defines if the class can expand
in any direction and thus have multiple objects in the instance.
\begin{verbatim}
data Expansion = None | Down | Right | Both
\end{verbatim}
There are several options:
\begin{itemize}
  \item the class can only have a single object, called a singleton, and thus
    does not expand, defined by the value \verb|None|;
  \item the class can have multiple objects that repeat downwards, defined by
    the value \verb|Down|;
  \item the class can have multiple objects that repeat to the right, defined by
    the value \verb|Right|; and,
  \item the class can have multiple objects that repeat both to the right and
    downwards, defined by the value \verb|Both|.
\end{itemize}

The layout is presented by a grid, where each cell in the Tabula grid
represents one or more cells in the spreadsheet. There are three kinds
of Tabula cells, represented by the type {\small\verb!TCell!}: \emph{input}, \emph{formula} and \emph{label}.

An \emph{input} cell is specified with the contents in the format
``name=default'', where \emph{name} is the name of the attribute it represents
and \emph{default} is the default value for the respective cells in the
spreadsheet. This value is also used to define the type of the contents, where a
numeric value implies the type being a number and a quoted text (possibly empty)
implies the type being textual.

A \emph{formula} cell is similar to an \emph{input} cell. It is defined using
the format ``name=formula'', where \emph{name} is the name of the attribute it
represents and \emph{formula} is the formula it defines. References in the
formula are made using attribute names defined by \emph{input} cells or by other
\emph{formula} cells.

A \emph{label} cell is any cell that is neither of kind \emph{input} nor
\emph{formula}. The content of the respective cells in the spreadsheet are
exactly the same content as in the Tabula cell. In this kind of cells fall empty
cells, numeric cells and any other textual cell.

The logic is as specified by the classes and the attributes defined in the grid.
Each class in the Tabula represents a kind of object present in the spreadsheet,
where the attributes are the ones in the grid contained in the class' range.

\paragraph{} ~\\[-1em]\indent
The spreadsheet of our running example 
can be defined with a Tabula containing a two-by-six grid with
two classes, where one class contains the list of items and another class
contains each item:
{\small
\begin{verbatim}
inventory =
  Tabula  "Inventory"
          -- classes
          [  Class "Inventory"  ((0,0), (1,5))  None
          ,  Class "Category"   ((0,2), (1,4))  Down
          ,  Class "Item"       ((0,3), (1,3))  Down]
          -- grid
          (Grid.fromLists  [  ["Inventory", ""]
                           ,  ["",          ""]
                           ,  ["desc=\"\"", "stock=0 : >=0"]
                           ,  ["",          ""]
                           ,  ["Total",     "total=SUM(stock)"]])
\end{verbatim}
}

A concise and graphical representation of this Tabula is presented in
Fig.~\ref{fig:example-inventory-tabula}. It describes the grid in its
tabular format and uses the background color to identify the classes
each cell belongs to\footnote{We assume that colors are available in
  the electronic version of this document,}. On the side there is a
legend relating the color to the class since the name is not always
present in the table: the class \class{Inventory} (in yellow) has a
label in its top-left cell but the classes \class{Category} and
\class{Item} (in green and blue, respectively) do not.  Moreover, in
some cases, the label can be different from the class name. The arrow
pointing down on the left of the attribute \attribute{desc} indicates
that the class \class{Item} expands down and the one on the left of
the left brace indicates that the class \class{Category} also expands
down, with the left brace removing any visual ambiguity that might
arise about what is repeated.

The layout of a Tabula can also expand horizontally  and
the expansion area can contain multiple columns or rows. This
configuration is displayed in
Fig.~\ref{fig:example-inventory1-years-tabula}, where the inventory
for one item (\textit{apple}) along the years is presented. Thus, a
new class \class{Year} is created, containing the attributes
\attribute{year}, \attribute{stock} and \attribute{sold} to represent
the year an object refers to, the stock of the item at the end of the
year and the number of items sold that year, respectively.

\begin{figure}[ht]
  \centering
  \small\sf
  %
  {
    \centering
    \begin{tabular}{|c|@{\hskip0pt}p{1\cmidrulesep}@{\hskip0pt}|l|r|r|r|r|p{2.8cm}|}
      \hhline{~~|------}
      \sstl   & & \sst{A}
        & \sst{B} & \sst{C}
        & \sst{D} & \sst{E}
        & \sst{F} \\\hhline{-~:======}
      \ssl{1} & & \textbf{Items}
        & 2012 &
        & 2013 &
        & \\\hhline{-~|------}
      \ssl{2} & &
        & \textbf{stock} & \textbf{sold}
        & \textbf{stock} & \textbf{sold}
        & \textbf{Average sold} \\\hhline{-~|------}
      \ssl{3} & &         apple
        & 5 &  12
        & 4 &  16
        & =AVERAGE(C3,E3) \\\hhline{-~|------}
    \end{tabular}
    \caption{Spreadsheet with an apple inventory per year.}
    \label{fig:example-inventory1-years-spreadsheet}
  }
  %
  {
    \centering
    \mbox{
    \begin{tabular}{|l||l|l||l|}
      \multicolumn{1}{l}{}
        & \multicolumn{2}{c}{\(\rightarrow\)}
        & \multicolumn{1}{l}{} \\\hhline{-||--||-}
      \cellcolor{classInventory}Inventory
        & \cellcolor{classYear}year=2000
        & \cellcolor{classYear}
        & \cellcolor{classInventory}
        \\\hhline{-||--||-}
      \cellcolor{classInventory}
        & \cellcolor{classYear}stock
        & \cellcolor{classYear}sold
        & \cellcolor{classInventory}Average sold
        \\\hhline{-||--||-}
      \cellcolor{classInventory}apple
        & \cellcolor{classYear}stock=0
        & \cellcolor{classYear}sold=0
        & \cellcolor{classInventory}avg=AVERAGE(sold)
        \\\hhline{-||--||-}
    \end{tabular}
    }
    \quad
    \mbox{
    \begin{tabular}{|c|l}
      \multicolumn{2}{c}{} \\
      \hhline{-~}
      \cellcolor{classInventory}~~ & Inventory \\
      \hhline{=~}
      \cellcolor{classYear}~~ & Year \\
      \hhline{-~}
    \end{tabular}
    }
    \caption{Tabula specification of an apple inventory per year.}
    \label{fig:example-inventory1-years-tabula}
  }
\end{figure}

%
In Fig.~\ref{fig:example-inventory-years-tabula}, the
conjugation of the configurations presented in Figs.~\ref{fig:example-inventory-tabula}
and~\ref{fig:example-inventory1-years-tabula} is shown, providing a more complex example. The result contains additional
classes that relate the expandable classes from both of the previous examples.
Thus, the
Tabula contains the class~\class{Inventory} from both examples, the
class~\class{Item} from Fig.~\ref{fig:example-inventory-tabula} and the class~\class{Year}
from Fig.~\ref{fig:example-inventory-years-tabula}. Moreover, it contains two new classes to
represent the attributes that are in the relation between:
\begin{itemize}
  \item \class{Category} and \class{Year}, named \class{CategoryYear}; and,
  \item \class{Item} and \class{CategoryYear}, named \class{ItemCategoryYear}.
\end{itemize}
Both of these classes expand in both directions: horizontally and vertically.  This Tabula is specified as follows:
{\small
\begin{verbatim}
inventoryYear =
  Tabula
    "InventoryYear"
    -- classes
    [  Class "Inventory"    ((0,0), (3,5))  None
    ,  Class "Year"         ((1,0), (2,5))  Right
    ,  Class "Category"     ((0,2), (3,4))  Down
    ,  Class "CategoryYear" ((1,2), (2,4))  Both
    ,  Class "Item"         ((0,3), (3,3))  Down
    ,  Class "ItemYear"     ((1,3), (2,3))  Both]
    -- grid
    (Grid.fromLists
      [["Inventory",  "year=2000",   "",           ""]
      ,["",           "",            "",           ""]
      ,["cat=\"\"",   "stock",       "sold",       "Average sold"]
      ,["desc=\"\"",  "stock=0:>=0", "sold=0:>=0", "avg=AVERAGE(sold)"]
      ,["",           "",            "",           ""]
      ,["Total",      "total=SUM(stock)", "",      ""]])
\end{verbatim}
}
or graphically as in Fig.~\ref{fig:example-inventory-years-tabula}.

\begin{figure}[ht]
  \centering
  \small\sf
  %
  {
    \centering
    \begin{tabular}{|c|@{\hskip0pt}p{1\cmidrulesep}@{\hskip0pt}|l|p{2.1cm}|r|p{2.1cm}|r|p{3.2cm}|}
      \hhline{~~|------}
      \sstl   & & \sst{A}
        & \sst{B} & \sst{C}
        & \sst{D} & \sst{E}
        & \sst{F} \\\hhline{-~:======}
      \ssl{1} & & \textbf{Inventory}
        & \hfill 2012 &
        & \hfill 2013 &
        & \\\hhline{-~|------}
      \ssl{2} & & & & & & & \\\hhline{-~|------}
      \ssl{3} & & Fruit
        & \cellLeftAligned{\textbf{stock}} & \cellLeftAligned{\textbf{sold}}
        & \cellLeftAligned{\textbf{stock}} & \cellLeftAligned{\textbf{sold}}
        & \cellLeftAligned{\textbf{Average sold}} \\\hhline{-~|------}
      \ssl{4} & & apple
        & \hfill 5 & 12
        & \hfill 4 & 16
        & =AVERAGE(C4,E4) \\\hhline{-~|------}
      \ssl{5} & & banana
        & \hfill 2 & 10
        & \hfill 3 & 12
        & =AVERAGE(C5,E5) \\\hhline{-~|------}
      \ssl{6} & & cherry
        & \hfill 8 & 9
        & \hfill 1 & 3
        & =AVERAGE(C6,E6) \\\hhline{-~|------}
      \ssl{7} & & & & & & & \\\hhline{-~|------}
      \ssl{8} & & Legumes~
        & \cellLeftAligned{\textbf{stock}} & \cellLeftAligned{\textbf{sold}}
        & \cellLeftAligned{\textbf{stock}} & \cellLeftAligned{\textbf{sold}}
        & \cellLeftAligned{\textbf{Average sold}} \\\hhline{-~|------}
      \ssl{9} & & beans
        & \hfill 7 & 5
        & \hfill 9 & 7
        & =AVERAGE(C9,E9) \\\hhline{-~|------}
      \ssl{10} & & peas
        & \hfill 10 & 10
        & \hfill 8 & 9
        & =AVERAGE(C10,E10) \\\hhline{-~|------}
      \ssl{11} & & & & & & & \\\hhline{-~|------}
      \ssl{12} & & \textbf{Total}
        & =SUM(B4:B6,\newline B9:B10) &
        & =SUM(D4:D6,\newline D9:D10) &
        & \\\hhline{-~|------}
    \end{tabular}
    \caption{Spreadsheet definition.}
    \label{fig:example-inventory-years-spreadsheet}
  }
  %
  {
    \centering
    \mbox{\small
    \begin{tabular}{c|l||l|l||l|}
      \multicolumn{1}{l}{}
        & \multicolumn{1}{l}{}
        & \multicolumn{2}{c}{\(\rightarrow\)}
        & \multicolumn{1}{l}{} \\\hhline{~-||--||-}
      & \cellcolor{classInventory}Inventory
        & \cellcolor{classYear}year=2000
        & \cellcolor{classYear}
        & \cellcolor{classInventory}
        \\\hhline{~-||--||-}
      & \cellcolor{classInventory}
        & \cellcolor{classYear}
        & \cellcolor{classYear}
        & \cellcolor{classInventory}
        \\\hhline{~=::==::=}
      \multirow{3}{*}{$\downarrow \left\{ \threelines \right. \downarrow$}
        & \cellcolor{green!30!white}cat=""
        & \cellcolor{cyan!20!white}stock
        & \cellcolor{cyan!20!white}sold
        & \cellcolor{green!30!white}Average sold
        \\\hhline{~=::==::=}
      & \cellcolor{blue!20!white}desc=""
        & \cellcolor{blue!30!white}stock=0 : >=0
        & \cellcolor{blue!30!white}sold=0 : >=0
        & \cellcolor{blue!20!white}avg=AVERAGE(sold)
        \\\hhline{~=::==::=}
      & \cellcolor{green!30!white}
        & \cellcolor{cyan!20!white}
        & \cellcolor{cyan!20!white}
        & \cellcolor{green!30!white}
        \\\hhline{~=::==::=}
      & \cellcolor{classInventory}Total
        & \cellcolor{classYear}total=SUM(stock)
        & \cellcolor{classYear}
        & \cellcolor{classInventory}
        \\\hhline{~-||--||-}
    \end{tabular}
    }
    \\
    \mbox{\small
    \begin{tabular}{|c|l}
      \multicolumn{2}{c}{} \\
      \hhline{-~}
      \cellcolor{yellow!30!white}~~ & Inventory \\
      \hhline{=~}
      \cellcolor{green!30!white}~~ & Category \\
      \hhline{-~}
    \end{tabular}
    }
    \mbox{\small
    \begin{tabular}{|c|l}
      \multicolumn{2}{c}{} \\
      \hhline{-~}
      \cellcolor{classYear}~~ & Year \\
      \hhline{=~}
      \cellcolor{cyan!20!white}~~ & CategoryYear \\
      \hhline{-~}
    \end{tabular}
    }
    \mbox{\small
    \begin{tabular}{|c|l}
      \multicolumn{2}{c}{} \\
      \hhline{-~}
      \cellcolor{blue!20!white}~~ & Item \\
      \hhline{=~}
      \cellcolor{blue!30!white}~~ & ItemCategoryYear \\
      \hhline{-~}
    \end{tabular}
    }
    \caption{Tabula specification of the spreadsheet.}
    \label{fig:example-inventory-years-tabula}
  }
\end{figure}

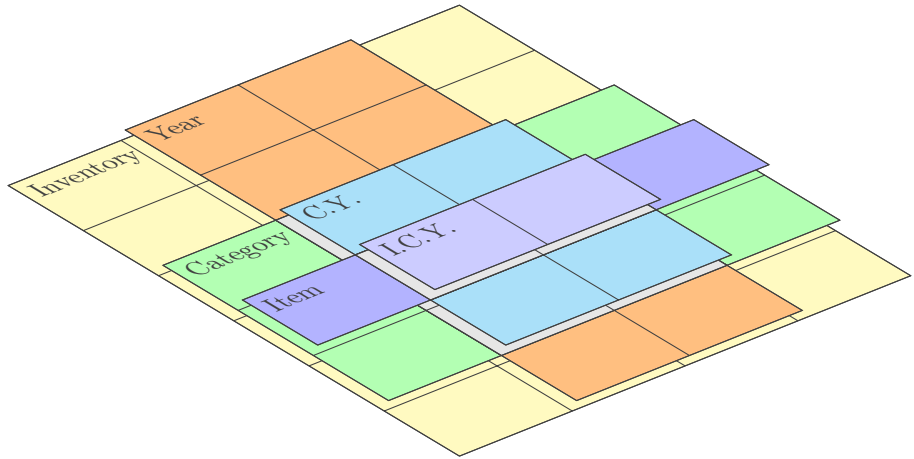
\begin{figure}[ht]
  \centering
  \begin{tikzpicture}
    \begin{scope}[draw=gray!50!black,text=gray!50!black,every node/.append style={yslant=0.4,xslant=-1},yslant=0.4,xslant=-1,y=-1cm,xscale=1.5]
      \draw[fill=yellow!30!white] (0, 0, -0.0) rectangle (4, 6, -0.0);
      \draw                       (0, 0)       grid      (4, 6);
      \draw[fill=green!30!white]  (0, 2, -0.3) rectangle (4, 5, -0.3);
      \draw[shift={(0,0,-0.3)}]   (0, 2)       grid      (4, 5);
      \draw[fill=classYear]       (1, 0, -0.3) rectangle (3, 6, -0.3);
      \draw[shift={(0,0,-0.3)}]   (1, 0)       grid      (3, 6);
      \draw[fill=gray!20!white]   (1, 2, -0.3) rectangle (3, 5, -0.3);
      \draw[shift={(0,0,-0.3)}]   (1, 2)       grid      (3, 5);
      \draw[fill=cyan!30!white]   (1, 2, -0.6) rectangle (3, 5, -0.6);
      \draw[shift={(0,0,-0.6)}]   (1, 2)       grid      (3, 5);
      \draw[fill=blue!30!white]   (0, 3, -0.6) rectangle (4, 4, -0.6);
      \draw[shift={(0,0,-0.6)}]   (0, 3)       grid      (4, 4);
      \draw[fill=gray!20!white]   (1, 3, -0.6) rectangle (3, 4, -0.6);
      \draw[shift={(0,0,-0.6)}]   (1, 3)       grid      (3, 4);
      \draw[fill=blue!20!white]   (1, 3, -0.9) rectangle (3, 4, -0.9);
      \draw[shift={(0,0,-0.9)}]   (1, 3)       grid      (3, 4);
      \node[below right] at (0, 0, -0.0) {\small Inventory};
      \node[below right] at (1, 0, -0.3) {\small Year};
      \node[below right] at (0, 2, -0.3) {\small Category};
      \node[below right] at (1, 2, -0.6) {\small C.Y.};
      \node[below right] at (0, 3, -0.6) {\small Item};
      \node[below right] at (1, 3, -0.9) {\small I.C.Y.};
    \end{scope}
  \end{tikzpicture}
  \caption{Class layers for the inventory example
  (Fig.~\ref{fig:example-inventory-years-tabula}).}
  \label{fig:example-inventory-years-layers}
\end{figure}

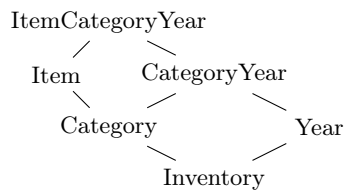
\begin{figure}[ht]
  \centering
  \begin{tikzpicture}[scale=0.7]
    \node (inv) at ( 0.0, 0.0) {\small Inventory};
    \node (cat) at (-2.0, 1.0) {\small Category};
    \node (yea) at ( 2.0, 1.0) {\small Year};
    \node (ite) at (-3.0, 2.0) {\small Item};
    \node (cy)  at ( 0.0, 2.0) {\small CategoryYear};
    \node (icy) at (-2.0, 3.0) {\small ItemCategoryYear};
    \draw (inv) -- (cat);
    \draw (inv) -- (yea);
    \draw (cat) -- (ite);
    \draw (cat) -- (cy);
    \draw (yea) -- (cy);
    \draw (ite) -- (icy);
    \draw (cy) -- (icy);
  \end{tikzpicture}
  \caption{Class ordering for the running example
  (Fig.~\ref{fig:example-inventory-years-tabula}).}
  \label{fig:classes-ordering}
\end{figure}

Classes are laid out in a layered fashion, where the class that is used by
another is at a higher level than the one that uses it. Thus, classes are not
broken down by other classes, but instead have parts of themselves covered by
other classes. This characteristic is demonstrated in
Fig.~\ref{fig:example-inventory-years-layers} with the ordering relation between
the classes show in Fig.~\ref{fig:classes-ordering}.

In order to have a valid model, some rules in laying the classes must be
followed:
\begin{itemize}
  \item There must always be a base class with the size of the Tabula grid. In
    the running example, this base class is \class{Inventory}.
  \item Inner classes must either use the whole height or the whole width of the
    class below. In the former case, the class is called a \emph{vertical
    class}, possibly expanding to the right, and in the latter case the class is
    called an \emph{horizontal class}, possibly expanding down. This rule
    prevents spreadsheet layout deformation when adding new instances of a
    class.
  \item An horizontal class must leave a line above and another below of the
    class below. A vertical class must leave a column to the left and another
    one to the right of the class below. This rule prevents ambiguity in the
    layer order and allows for a simpler visual representation. Note that there
    is no need to leave space between two classes at the same level.
  \item Two classes are allowed to intersect when one is contained by the other,
    i.e., the intersection area is the area of the contained class, or when a
    class is a vertical one and the other is an horizontal class.
  \item When a vertical class intersects with an horizontal one, a relation
    class must use the intersection area.
  \item Only relation classes can expand both to the right and down at once.
\end{itemize}
Note that this rules have a similar objective as the tiling rules for
ClassSheets.

The notion of layers and class ordering is also used to define which class an
attribute belongs to. Thus, an attribute belongs to the top-most class over that
cell. The name resolution and translation to cell references in the instance
works like in ClassSheets.

\subsection{Model-Driven Spreadsheet Evolution}

In a model-driven spreadsheet development environment, spreadsheet
users can reason about large and complex data just by looking at the
model. In this setting, spreadsheet users are not only able to
edit/transform the instance (as in regular spreadsheet systems), but
also to edit/transform the model. However, after a transformation in
one of the software artifacts (model or instance), the model-driven
environment has to guarantee their synchronization. That is to say,
the instance has always to conform to the model, after every
instance or model transformations. 

In our previous work, model/instance co-evolution was applied to
ClassSheets with a bidirectional transformation
engine~\cite{icmt2012,conf/icse/CunhaFMS12} that is able to perform a
change in the ClassSheet and then have the respective spreadsheet
co-evolved correspondingly.  Moreover, this transformation engine is
also able to perform changes in the spreadsheet that possibly breaks
conformance with the model, but that co-evolves the model in order to
keep the conformance, and therefore the bidirectional capability.

\begin{figure}[ht]
  \centering
  \includegraphics[scale=0.75]{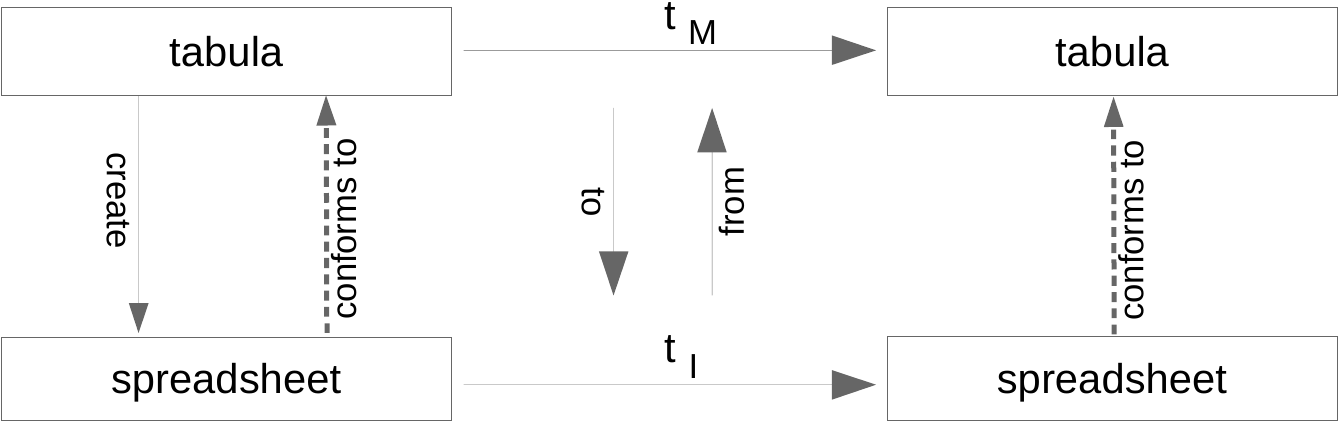}
  \caption{Diagram of the bidirectional transformation engine.}
  \label{fig:bx}
\end{figure}

We have adapted and improved such bidirectional transformation engine
to guarantee synchronization after a Tabula or a spreadsheet instance
evolution.  The structure of the bidirectional transformation engine
is shown in Fig.~\ref{fig:bx}: the engine includes a set of
transformation operations both on the model $t_M$ and on the instance
$t_I$. It also defines two mapping functions ($to$ and $from$) between
  such operations, that convert transformations in the model to
  transformation in the instance and \textit{vice versa}. These two
  functions guarantee the synchronization of model and instance: An
  operation on the model (instance) is transformed to operations in
  the instance (model), such that after applying the transformations
  in the model and instance the conformance relation is guaranteed.

The arrow $create$ from a Tabula model to a spreadsheet instance means
that from the model an initial instance can be generated, where the
model serves as a type system. Thus, when editing the generated instance,
end users are forced to conform to the model. That is to say that they
have to define spreadsheet data/value with that specific Tabula type.

In this paper we omit the definition of this bidirectional
engine. Such definition is provided in~\cite{phdJorgeMendes}.

\section{Tabula With Real-World Spreadsheets}
\label{sec:eval}
In order to assess the expressiveness of Tabula, when compared to the
ClassSheet model, let us consider a widely used spreadsheet for a
personal budget provided by Microsoft%
\footnote{\emph{Basic personal budget} spreadsheet
available at:\\\url{https://templates.office.com/en-us/Basic-personal-budget-TM16400272}}.

The Microsoft budget spreadsheet template has 14 columns and 91 rows,
which users can reuse and edit by adding more data (rows) to the
spreadsheet. Column-wise, the first one contains labels for each row,
the twelve next columns contain the data for the twelve months of a
year, and the last one contains the total for the year for each
item. Row-wise, the spreadsheet has a title, then a header for the
columns, a set of rows for the income, and then sets of rows for the
expenses, grouped into categories (\textit{home}, \textit{daily
living},
\textit{transportation}, \textit{entertainment}, \textit{health},
\textit{vacations}, \textit{recreation}, \textit{dues/subscriptions},
\textit{personal}, \textit{financial obligations} and \textit{misc. payments}).
The categories have 5 items in average.

\begin{figure}[htb!]
  \centering
  \includegraphics[scale=0.30]{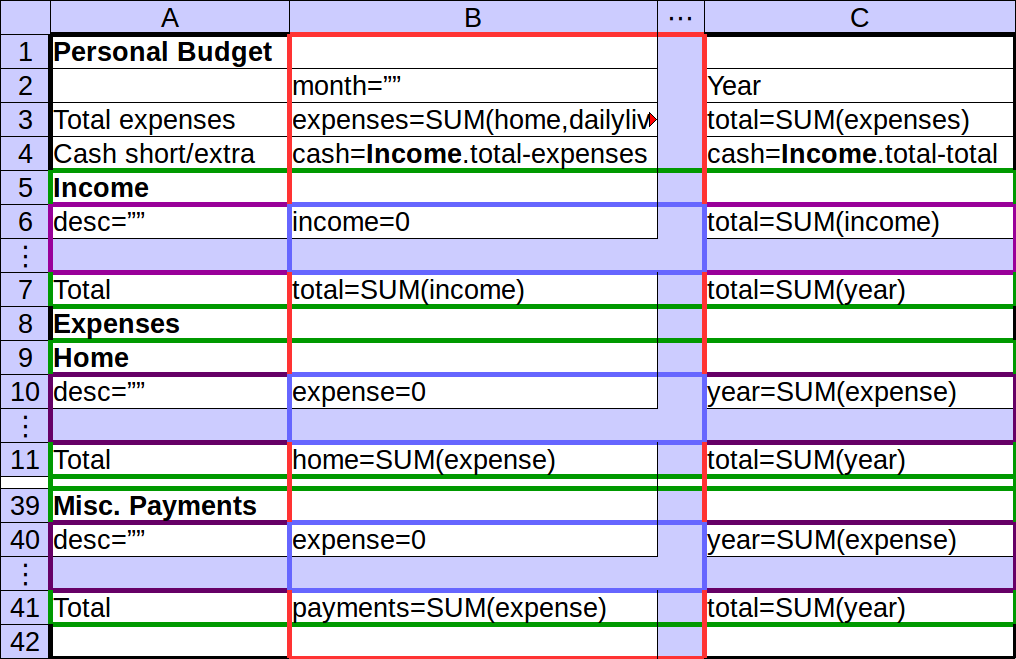}
  \caption{A ClassSheet for the Microsoft personal budget spreadsheet, with rows 12--38 omitted.}
  \label{fig:personal-budget-classsheet}
\end{figure}

In fact, this spreadsheet was used in an empirical
study~\cite{tse2015} to evaluate the performance of spreadsheet users
when using model-driven spreadsheets. In that study the business logic
of the budget spreadsheet was modeled by a ClassSheet. ClassSheets,
however, could not capture the essence of the spreadsheet, resulting
in a large and hard to understand and maintain model. A fragment of
this ClassSheet model is shown in
Fig.~\ref{fig:personal-budget-classsheet}.

Because nested repetitions are not supported by ClassSheets, the
model of the spreadsheet has to contain the concrete type of expenses
(11 in total). This result in a visual ClassSheet that has 14 columns
and 42 rows, Moreover, the formulas have also to refer to a large
number of unique attributes, making their definition more complex and
prone to errors. Finally, users cannot extend the categories of the
expenses in the spreadsheet, only. A change in the model is needed,
too.

In comparison, the Tabula model presented in
Fig.~\ref{fig:personal-budget-tabula} for the same spreadsheet can
model the different kinds of expenses and thus be more concise than
the ClassSheet model. Moreover, there are no explicit reference to
specific type of expenses (like, for example, \textit{Home}
and \textit{MiscPayments} in the ClassSheet model). Finally, the
Tabula also define constraints on types, like specification that only
positive numbers can be added as expenses. The Tabula has the same
number of columns (14), as the template is not dynamic in that axis,
but has only 12 rows, describing the kinds of expenses in one
expandable class.

\begin{figure}[htb!]
  \centering
  %
  \colorlet{classBudget}{yellow!25}
  \colorlet{classMonth}{cyan!25}
  \colorlet{classIncome}{green!30}
  \colorlet{classIncomeItem}{green!50}
  \colorlet{classIncomeMonth}{NavyBlue!50}
  \colorlet{classIncomeItemMonth}{NavyBlue!75}
  \colorlet{classExpense}{orange!50}
  \colorlet{classExpenseItem}{orange}
  \colorlet{classExpenseMonth}{violet!25}
  \colorlet{classExpenseItemMonth}{violet!50}
  {
    \centering
    \small\sf
    \mbox{
    \begin{tabular}{r|l||l||l|}
      \multicolumn{1}{l}{}
        & \multicolumn{1}{l}{}
        & \multicolumn{1}{c}{\(\rightarrow\)}
        & \multicolumn{1}{l}{}
        \\\hhline{~-||-||-}
      & \cellcolor{classBudget}Personal Budget
        & \cellcolor{classMonth}
        & \cellcolor{classBudget}
        \\\hhline{~-||-||-}
      & \cellcolor{classBudget}
        & \cellcolor{classMonth}month=""
        & \cellcolor{classBudget}Year
        \\\hhline{~-||-||-}
      & \cellcolor{classBudget}Total Expenses
        & \cellcolor{classMonth}total=SUM(Expense.total)
        & \cellcolor{classBudget}total=SUM(Expense.total)
        \\\hhline{~-||-||-}
      & \cellcolor{classBudget}Cash short/extra
        & \cellcolor{classMonth}cash=Income.total-total
        & \cellcolor{classBudget}cash=Income.total-total
        \\\hhline{~-||-||-}
      & \cellcolor{classIncome}Income
        & \cellcolor{classIncomeMonth}
        & \cellcolor{classIncome}
        \\\hhline{~=::=::=}
      \(\downarrow\) & \cellcolor{classIncomeItem}desc=""
        & \cellcolor{classIncomeItemMonth}income=0 : >=0
        & \cellcolor{classIncomeItem}year=SUM(income)
        \\\hhline{~=::=::=}
      & \cellcolor{classIncome}Total
        & \cellcolor{classIncomeMonth}total=SUM(income)
        & \cellcolor{classIncome}total=SUM(year)
        \\\hhline{~-||-||-}
      & \cellcolor{classBudget}Expenses
        & \cellcolor{classMonth}
        & \cellcolor{classBudget}
        \\\hhline{~=::=::=}
      \multirow{3}{*}{\(\downarrow \left\{ \threelines \right. \downarrow\)}
        & \cellcolor{classExpense}cat=""
        & \cellcolor{classExpenseMonth}stock
        & \cellcolor{classExpense}Average sold
        \\\hhline{~=::=::=}
      & \cellcolor{classExpenseItem}desc=""
        & \cellcolor{classExpenseItemMonth}stock=0 : >=0
        & \cellcolor{classExpenseItem}avg=AVERAGE(sold)
        \\\hhline{~=::=::=}
      & \cellcolor{classExpense}Total
        & \cellcolor{classExpenseMonth}total=SUM(expense)
        & \cellcolor{classExpense}total=SUM(year)
        \\\hhline{~=::=::=}
      & \cellcolor{classBudget}
        & \cellcolor{classMonth}
        & \cellcolor{classBudget}
        \\\hhline{~-||-||-}
    \end{tabular}
    }
    \\
    \mbox{
    \begin{tabular}{|c|l}
      \multicolumn{2}{c}{} \\
      \hhline{-~}
      \cellcolor{classBudget}~~ & Budget \\
      \hhline{=~}
      \cellcolor{classMonth}~~ & Month \\
      \hhline{-~}
    \end{tabular}
    }
    \mbox{
    \begin{tabular}{|c|l}
      \multicolumn{2}{c}{} \\
      \hhline{-~}
      \cellcolor{classIncome}~~ & Income \\
      \hhline{=~}
      \cellcolor{classIncomeItem}~~ & IncomeItem \\
      \hhline{=~}
      \cellcolor{classIncomeMonth}~~ & IncomeMonth \\
      \hhline{=~}
      \cellcolor{classIncomeItemMonth}~~ & IncomeItemMonth \\
      \hhline{-~}
    \end{tabular}
    }
    \mbox{
    \begin{tabular}{|c|l}
      \multicolumn{2}{c}{} \\
      \hhline{-~}
      \cellcolor{classExpense}~~ & Expense \\
      \hhline{=~}
      \cellcolor{classExpenseItem}~~ & ExpenseItem \\
      \hhline{=~}
      \cellcolor{classExpenseMonth}~~ & ExpenseMonth \\
      \hhline{=~}
      \cellcolor{classExpenseItemMonth}~~ & ExpenseItemMonth \\
      \hhline{-~}
    \end{tabular}
    }
  }
  \caption{A Tabula for the Microsoft personal budget spreadsheet.}
  \label{fig:personal-budget-tabula}
\end{figure}

A summary of the differences between both models is presented in
Table~\ref{tab:metrics}, which for each entry contains the number of
rows, columns, classes, attributes, input cells and formulas.

\begin{table}
  \centering
  \caption{Comparison of different metrics for the original personal
  spreadsheet, its ClassSheet and its Tabula.}
  \label{tab:metrics}
  \setlength{\tabcolsep}{7pt}
  \begin{tabular}{lrrrrrr}
    \cline{2-7}
                      & width & height & classes & attributes & input & formulas \\\hline
    personal budget   &    14 &     91 &  \cc{-} &        988 &   744 &       244 \\[0.2em]
    Tabula            &    14 &     12 &       5 &         81 &    27 &        54 \\
    ClassSheet        &    14 &     42 &      25 &        350 &   156 &       194 \\[0.2em]
    Tabula (dyn.)     &     3 &     12 &      10 &         16 &     6 &        10 \\
    ClassSheet (dyn.) &     3 &     42 &      38 &         65 &    25 &        40 \\\hline
  \end{tabular}
\end{table}

The first entry in this table is the \emph{personal budget} (as
obtained from the link provided above). The second and fourth entry
contains two Tabula models for this personal budget where the first
model specifies the twelve months, and the second the months are
defined via a repetition (this is the Tabula in
Fig.~\ref{fig:personal-budget-tabula}). The third and fifth entries
contains the metrics of the ClassSheet models for the personal budget
(with the twelve months defined or with repetitions as shown in
Fig.~\ref{fig:personal-budget-classsheet}). As we can see in this table
Tabula is 20\%--26\% smaller than the equivalent ClassSheet in terms of the
number of classes and attributes.

\section{Conclusion}
\label{sec:conclusion}
In this paper we present the tabula language to model spreadsheet
tables. Tabula was inspired by previous work on model-driven
engineering, namely by the (visual) ClassSheet language, but it
provides more flexibility and expressive power to model real-world
spreadsheets. 

We have modeled a real personal budget spreadsheet, and we have
compared the ClassSheet and Tabula models. Our preliminary results
show that Tabula is able to model such spreadsheet in a natural way,
as opposed to ClassSheets.

Tabula defines a kind of type system for spreadsheets. As future work
we intend to extend this type language such that it is possible to
specify more precisely the type of values a spreadsheet cell may
contain (for example, a cell may contain units like \textit{seconds},
\textit{inches}, etc.). This will prevent a larger set of errors in
spreadsheets, e.g., by preventing operations with incompatible units.

\subsubsection{Acknowledgments.} 

This work is financed by the ERDF – European Regional Development Fund
through the Operational Programme for Competitiveness and
Internationalisation - COMPETE 2020 Programme and by National Funds
through the Portuguese funding agency, FCT - Fundação para a Ciência e
a Tecnologia within project POCI-01-0145-FEDER-016718, by the
bilateral project FCT/DAAD with ref.~441.00, and by a PhD scholarship
from FCT with ref.~SFRH/BD/112651/2015.

\bibliographystyle{splncs03}
\bibliography{references}

\end{document}